\documentclass[reprint,twocolumn,showpacs,nofootinbib,prd,amssymb]{revtex4}

\usepackage[usenames,dvipsnames]{xcolor}

\usepackage{amssymb, amsmath}
\usepackage{epsf}
\usepackage{bm}
\usepackage{ulem}

\usepackage{natbib}
\usepackage{hyperref}
\usepackage{graphicx}
\usepackage{cancel}

\newcommand{\bs}{\boldsymbol}
\newcommand{\p}{\partial}

\newcommand{\divv}{\text{div}}

\newcommand{\const}{\text{const}}

\newcommand{\aap}{Astron.\ Astrophys.}
\newcommand{\mnras}{Mon.\ Not.\ R.\ Astron.\ Soc.}

\newcommand{\apjl}{Astrophys.\ J.\ Lett.}

\def\la{\; \raise0.3ex\hbox{$<$\kern-0.75em\raise-1.1ex\hbox{$\sim$}}\;}
\def\ga{\;  \raise0.3ex\hbox{$>$\kern-0.75em\raise-1.1ex\hbox{$\sim$}}\;}

\begin{document}

	\title{Hyperon bulk viscosity effects in neutron-star inspirals}
	%
	\author{Elena M. Kantor, Mikhail E. Gusakov, Kirill Y. Kraav}
	\affiliation{
		Ioffe Institute,
		Politekhnicheskaya 26, 194021 Saint-Petersburg, Russia
	}
	\date{}
	

	\begin{abstract}
		
	The paper revisits the role of hyperon bulk viscosity during neutron-star inspiral. We find that hyperon bulk viscosity exerts only a minor influence on the gravitational-wave phase, causing a phase shift of $\sim 10^{-3}\,\rm rad$. This shift is too small to be detected by existing gravitational-wave observatories, and it seems unlikely that next-generation detectors will be able to isolate this effect from other factors. However, our analysis indicates that hyperon bulk viscosity can significantly heat the hyperon core of a neutron star, raising its temperature to approximately $5 \times 10^8\,\rm K$. 
	
	\end{abstract}

	\maketitle

\section{Introduction}

Neutron stars (NS) are extraordinary celestial objects containing degenerate matter with densities exceeding those of atomic nuclei. Since such thermodynamically cold and extremely dense matter currently cannot be reproduced in terrestrial laboratories, looking into NS observational manifestations offers a unique opportunity to constrain and explore its physical properties. A particularly promising option is to study the gravitational-wave (GW) signals originating from the binary NS inspirals and mergers, which have been observed since the first detection, GW170817, reported in 2017 by the LIGO and Virgo collaborations \cite{GW17}. A reliable interpretation of such observations requires a detailed understanding of the physics behind the formation of the GW signal, in particular, the way it encodes the properties of the stellar matter.

The simplest approximation to model the GW signal from inspiraling stars is to treat the latter as point-mass objects, thereby ignoring their internal degrees of freedom. This approximation, however, is not suitable for exploring the properties of the stellar matter, as within the point-mass model the emitted GW signal 
is insensitive to the equation of state/internal NS microphysics.
A more detailed treatment of the problem requires relaxing the point-mass approximation and accounting for  finite-size effects. The dominant correction to the point-mass model, in particular, comes from accounting for the deformations of a star under the gravitational tidal forces imposed by its companion. These tidally induced deformations enhance the GW radiation from the binary, causing the inspiral to accelerate and, therefore, modifying the associated GW signal \cite{fh08}. Accounting for this effect when interpreting the GW170817 signal has already allowed to put constraints on the NS tidal deformability (see, e.g., Ref.\ \cite{constraint19}). While numerous weaker finite size effects (see, e.g., the review \cite{dhs21}) that also take place are likely beyond the reach of the current detectors, they, hopefully, might be resolved by future detectors with enhanced sensitivity and broader frequency ranges, such as the Einstein Telescope and Cosmic Explorer.
Therefore, precise interpretation of the forthcoming observations requires accurate models of the inspiraling neutron stars, incorporating the influence of the various finite size effects on the GW signal.

Significant progress has already been made in this area
(see, e.g., the reviews \cite{dhs21,sedrakian26} and references therein).
One of the challenges is to model how energy dissipation inside a star affects the GW signal. The basic idea is that part of the binary orbital energy is continuously spent on maintaining the tidally induced stellar deformation in the presence of dissipative losses. This dissipation heats the star and accelerates the inspiral, thereby modifying the associated GW signal. At high temperatures, the dominant dissipative mechanism is bulk viscosity, which is associated with energy losses caused by deviations of the matter from chemical equilibrium.
In stars with standard nucleonic composition, the relevant chemical reactions are weak leptonic Urca processes. If the matter contains hyperons, however, the dominant contribution to the bulk viscosity may come from weak nonleptonic reactions \cite{oghf19}. These reactions proceed much faster than Urca processes and can therefore lead to substantially enhanced bulk-viscous dissipation in hyperonic matter. A similar mechanism operates in strange quark matter, where bulk viscosity is enhanced by weak nonleptonic processes involving strange quarks \cite{quark1,quark2}. Thus, modeling the effect of bulk viscosity on the stellar temperature and GW signal under different microphysical assumptions, and comparing the results with observations, may provide a probe of hyperons or strange quark matter in NS interiors.

Some existing studies of bulk viscosity in NS inspirals have reached apparently conflicting conclusions. In particular, studies of nucleonic bulk viscosity \cite{aw19,most22,ray23} and of bulk viscosity produced by weak nonleptonic reactions \cite{gpkd24,debarati25a,debarati25b} differ in their assessment of its astrophysical importance. On the one hand, Ref.\ \cite{most22} suggests that even bulk viscosity as strong as the hyperonic one is likely to have only a weak effect on the inspiral. On the other hand, Refs.\ \cite{gpkd24,debarati25a,debarati25b} find that bulk viscosity produced by weak nonleptonic reactions may lead to significant stellar heating and potentially detectable GW phase shifts. This discrepancy motivates us to revisit the role of bulk viscosity during NS inspirals. Section~\ref{theory} introduces the theoretical framework used in our analysis. We present our numerical results in Sec.~\ref{num} and summarize our findings in Sec.~\ref{disc}.

\section{Theoretical framework}\label{theory}
\subsection{Model of a neutron star with hyperonic composition}

In this study, we use a simplified fluid model of a neutron star. The stellar interior is divided into two regions: an outer barotropic layer, representing the crust, and an inner nonbarotropic region, representing the core. We neglect the elasticity of the crust and treat the entire star as a degenerate, locally charge-neutral fluid. This approximation is sufficient for our present purpose, which is to estimate the influence of bulk-viscous dissipation in the core on the thermal and orbital evolution during the inspiral.

The particle composition of the core is model-dependent and varies with density. Here, we assume that, as predicted by a number of modern equations of state (e.g., \cite{ghk14, rsw2018, negreirosetal2018, fortinetal2017, providenciaetal2019}), the core matter is composed of neutrons $(n)$, protons $(p)$, electrons $(e)$, and, at higher densities, muons $(\mu)$, $\Lambda$ hyperons $(\Lambda)$, and $\Xi^-$ hyperons $(\Xi^-)$. Microscopic calculations indicate that $\Xi^-$ hyperons may be strongly paired throughout the density range in which they appear under neutron-star conditions; see, e.g., \cite{sc19}. Since $\Xi^-$ hyperons are charged, we refer to this state as superconducting. The pairing state of $\Lambda$ hyperons is less certain. 
Available calculations based on the results of Ref.\ \cite{takahashietal2001} suggest, however, that the $\Lambda$-hyperon pairing gap may be substantially smaller than the $\Xi^-$ gap \cite{ws2010, takatsukaetal2006}.
We therefore treat $\Lambda$ hyperons as normal. Finally, neutron and proton pairing gaps may also vanish (or be reasonably small) in the density and temperature ranges relevant to our problem \cite{sc19}. Thus, for simplicity, in what follows, we assume that all particle species except $\Xi^-$ hyperons are normal, whereas $\Xi^-$ hyperons are strongly superconducting. We also examine the possible effects of 
nucleon superfluidity
on our results below in Sec.~\ref{num}.

The core matter of an NS is subject to numerous reactions of particle mutual transformations. First of all, similar to nucleonic matter, various weak leptonic (Urca) reactions are present. They, however, can be safely ignored as slow and inefficient at temperatures and perturbation frequencies relevant for NS inspirals%
\footnote{We note in this connection that it has been argued in the literature
(see, e.g., Ref.~\cite{aw19}) that neutron stars with standard nucleonic
composition may be heated
during the inspiral up to $2\times 10^8\,{\rm K}$ by Urca reactions.
According to Ref.~\cite{aw19}, the dominant contribution to this heating comes
from resonantly excited g-modes. This estimate should, however, be treated with
some caution. The excitation of g-modes is controlled by the corresponding
overlap integrals, i.e.\ by the scalar products of the mode eigenfunctions with
the tidal-force field. These overlap integrals are especially sensitive to the
consistency of the perturbation calculation, because g-modes are nearly
orthogonal to the tidal potential.
In Ref.~\cite{aw19}, relativistic background neutron-star models were used,
whereas the eigenmodes were computed using Newtonian oscillation equations.
In addition, metric factors were not included in the expression for the overlap
integral; see their equation~(17). Such a mixed treatment may noticeably enhance
the g-mode overlap integrals. The reason is that
even small errors in the eigenfunctions may spoil the near-orthogonality of
g-modes to the tidal potential and thereby produce an enhanced overlap. A
similar issue for interface modes was discussed in Ref.~\cite{pap21}; see the
penultimate paragraph of their section~4.2.}.
In addition to Urca processes the hyperonic matter may also support the so-called weak nonleptonic processes, which are much faster and, generally, should be accounted for. 
For the specific $npe\mu\Lambda\Xi^-$ composition considered here, the relevant weak nonleptonic reactions are \cite{oghf19}:
\begin{gather}
	n+p \leftrightarrow \Lambda +p, \label{reac1} \\
	n+n \leftrightarrow \Lambda +n,  \\
	n+\Lambda \leftrightarrow \Lambda +\Lambda,  \label{reac3}\\
	n+\Xi^- \leftrightarrow \Lambda +\Xi^-,  \label{reac4}\\
	n+\Lambda \leftrightarrow \Xi^- +p. \label{reac5}
\end{gather}
Because $\Xi^-$ hyperons are assumed to be strongly superconducting, the weak nonleptonic reactions involving them are exponentially suppressed. We therefore treat reactions (\ref{reac4}) and (\ref{reac5}) as frozen. If both $\Lambda$ and $\Xi^-$ hyperons are present, the following strong reaction may also occur:
\begin{gather}
	2\Lambda \leftrightarrow \Xi^- +p.   \label{strong}
\end{gather}
In nonsuperfluid and nonsuperconducting matter, reaction~(\ref{strong}) is much faster than the weak nonleptonic reactions: its rate exceeds typical weak rates by about $14$--$16$ orders of magnitude \cite{oghf19,kgk25}. The superconductivity of $\Xi^-$ hyperons, however, suppresses this rate. Repeating the estimate of section~IID of Ref.~\cite{kgk25}, with proton superconductivity replaced by $\Xi^-$ superconductivity, one may write the suppression factor in the form
\begin{gather}
R_\Xi \simeq 0.221\,\tau_\Xi^{-3.5}
\exp\left(-\frac{1.764}{\tau_\Xi}\right),
\qquad
\tau_\Xi \equiv \frac{T}{T_{c\Xi}},
\end{gather}
where $T_{c\Xi}$ is the critical temperature for the onset of $\Xi^-$ superconductivity and $T$ is temperature.
This estimate shows that, for sufficiently small $\tau_\Xi$ (roughly $\tau_\Xi \lesssim 0.04$), the strong process may become comparable to, or even slower than, the weak nonleptonic reactions. Thus, whether reaction~\eqref{strong} is fully equilibrated on the timescale of tidal perturbations is model-dependent and cannot be established without specifying the poorly known $\Xi^-$ pairing gap. At the same time, as discussed in section~IID of Ref.~\cite{kgk25}, treating this strong process as frozen or equilibrated is not expected to substantially affect the final results: the corresponding maximum bulk viscosity is smaller than that produced by the weak nonleptonic reactions, while the associated changes in the adiabatic index are also expected to be modest. In what follows, we therefore choose the equilibrated limit for definiteness and impose chemical equilibrium with respect to reaction~\eqref{strong}.

\subsection{Perturbation equations of a neutron star in a binary system}
\label{perturb} 

To begin with, let us introduce the key notations and conventions to be used in our calculations. First of all, we characterize the stellar matter by the pressure $P$, energy density $\varepsilon$, enthalpy density $w\equiv P+\varepsilon$ and temperature $T$, and each particle species $k=(n,p,e,\mu,\Lambda, \Xi^-)$ by the corresponding number densities $\{n_k\}$, chemical potentials $\{\mu_k\}$, and electric charges $\{e_k\}$. Second, to describe the motion of the stellar matter, we introduce the collective hydrodynamic 4-velocity $u^\mu=(u^t,\bs{u})$, shared by all particle species. 
Such a description is justified regardless of the assumed superconductivity of
$\Xi^-$ hyperons, since local charge neutrality electromagnetically locks
$\Xi^-$ hyperons to the other charged particle species and suppresses their
relative motion (see, e.g., \cite{gkcg13, dg16, kgk2024} for similar arguments
in superconducting nucleonic matter).
Finally, we describe the gravitational field using the metric tensor $g_{\mu\nu}$ with $\{-,+,+,+\}$-signature. Here and in what follows, we use Latin indices $(i,j,k,\dots)$ to label different particle species, and Greek indices $(\mu,\nu,\lambda,\dots)$ to denote tensor components. Unless stated otherwise, we always imply summation over repeated Greek and Latin indices. The bold font is used for 3-vectors, such as, e.g., the spatial component $\bs{u}$ of the 4-velocity $u^\mu$. For convenience, we also employ natural units where the speed of light $c=1$ and Boltzmann constant $k_{\rm B}=1$.

In this study, we consider perturbations of a nonrotating neutron star in a
binary system caused by the gravitational field of its companion. To this end,
we introduce Eulerian perturbations, $\delta f=f-f_0$, which describe the
deviation of a quantity $f$ from its value $f_0$ in an isolated neutron star in
chemical and hydrostatic equilibrium. For simplicity, we work in the Cowling
approximation, i.e., we neglect 
the effect of stellar perturbations on the gravitational field.
In this approximation, the deviation of the gravitational field from
that of an isolated neutron star is due solely to the external gravitational
field of the companion.

At large binary separations, the companion's gravitational field in the
vicinity of the considered star can be approximately described by the
companion's gravitational potential $U$. It is customary to decompose this
potential as $U=U_0+\tilde U$. The first term, $U_0$, produces the same
acceleration at every point inside the star. It may therefore affect the motion
of the star as a whole, but it cannot deform the star. Stellar deformations are
caused only by the nonuniform part of the external field, described by the
tidal potential $\tilde U$. The explicit form of these potentials will be given
in Sec.~\ref{sec24}. Here it is sufficient to note that, at the large binary
separations considered below, the external field is weak compared with the
star's own gravitational field and can therefore be included perturbatively, as
discussed in Sec.~\ref{sec24}.

To describe tidal perturbations, we follow \cite{kgk24} and employ a reference frame whose origin follows the stellar center of mass, accelerated solely by potential $U_0$. In this frame, within approximations made, the interval $ds^2=g_{\mu\nu}dx^\mu dx^\nu$ can be written in spherical coordinates $x^\mu=(t,r,\theta,\varphi)$ as
\begin{gather}
	\label{gtensor}
	ds^2=-e^{\nu_0+2\tilde{U}}dt^2+e^{\lambda_0}dr^2+r^2(d\theta^2+\sin^2\theta d\varphi^2),
\end{gather}
where $\nu_0(r)$ and $\lambda_0(r)$ are the standard metric functions that describe the equilibrium spacetime of an isolated nonrotating neutron star. We also introduce the so-called Lagrangian displacement $\xi^\mu=(\xi^t,\bs{\xi})$ that shows the perturbations of the fluid element worldlines and relates to the velocity perturbation in this frame by the equality 
$\bs u=e^{-\nu_0/2}\dot{\bs\xi}$, 
where we used the abbreviated notation $\dot{f}\equiv\p f/\p t$ for the time-derivative. 

Within approximations made, small perturbations of a neutron star from its equilibrium state at $U=0$ in the chosen reference frame are governed by the following equations:
\begin{gather}
		e^{-\nu_0}w_0\ddot{\bs\xi}=-{\pmb\nabla}\delta P+\frac{\delta w}{w_0}{\pmb\nabla} P_0-w_0{\pmb\nabla} \tilde{U} 	\label{eu}\\
		\delta \dot{n}_k+\divv(n_{k0}{\dot{\bs\xi}})=e^{\nu_0/2}\Delta \Gamma_k, \label{cont}
		\\
        \label{thermo1}
        \delta w=\delta P + \delta\varepsilon, \quad \delta \varepsilon=\mu_{k0}\delta n_k, \quad \delta P=n_{k0}\delta\mu_k, \\
        \delta\mu_k=\biggl(\frac{\p\mu_k}{\p n_i}\biggr)_0\delta n_i, \quad \delta P=\biggl(\frac{\p P}{\p n_k}\biggr)_0\delta n_k.
        \label{deltaPmu}
\end{gather}
Here ${\pmb\nabla}$ and $\divv$ are the three-dimensional covariant operators
associated with the unperturbed spatial metric $dl^2=e^{\lambda_0}dr^2+r^2(d\theta^2+\sin^2\theta d\varphi^2)$; see Appendix B of \cite{kgk24} for details.
Equation~\eqref{eu} represents the linearized Euler equation, obtained from the
spatial projection of the energy-momentum conservation law,
\begin{gather}
    \delta\left[
    \left(g^{\rho\nu}+u^\rho u^\nu\right)
    \nabla_\mu T^\mu_{\ \nu}
    \right]=0 ,
\end{gather}
where
\begin{gather}
\label{perfectTmunu}
    T^{\mu\nu}\equiv w u^\mu u^\nu + P g^{\mu\nu}
\end{gather}
is the perfect-fluid stress-energy tensor, and $\nabla_\mu$ is the covariant
derivative associated with the metric \eqref{gtensor}.
Furthermore, Eq.~\eqref{cont} represents the set of linearized continuity
equations for the different particle species. The source terms
$\Delta\Gamma_k$ account for particle conversion in chemical reactions,
specifically in the weak nonleptonic processes
\eqref{reac1}--\eqref{reac3} and in the strong process \eqref{strong} in the
hyperonic core. The remaining relations in the third and fourth lines follow
from the thermodynamic identities
\begin{gather}
    d\varepsilon=\mu_k dn_k, 
    \quad 
    dP=n_k d\mu_k, 
    \quad 
    w\equiv P+\varepsilon=\mu_k n_k,
    \label{tdrel}
\end{gather}
and from the equation of state (EOS), which is assumed to specify the energy
density $\varepsilon=\varepsilon(\{n_k\})$ as a function of the number densities
$\{n_k\}$.

Physical perturbations described by the above equations must be regular at the
stellar center and satisfy the condition that the total pressure vanish at the
perturbed stellar surface. These are the boundary conditions imposed in our
calculations.

\subsection{Energy dissipation rate}

Tidal perturbations induced by a companion star drive the stellar matter out
of chemical equilibrium and thereby dissipate the energy of the perturbation.
This has two consequences. First, dissipation heats the star. Second, it
accelerates the inspiral, because part of the orbital energy is transferred to
tidally induced perturbations and then dissipated. To quantify these effects,
we need the energy dissipation rate $\dot{E}_{\rm diss}$ associated with
chemical reactions. This rate can be obtained from the energy balance equation
for the perturbed star.

To derive the energy 
balance equation,
we note that the Euler equation \eqref{eu} can be equivalently rewritten as 
\begin{gather}
e^{-\nu_0/2}w_0 \ddot{\bs\xi}=-n_{i0}{\pmb\nabla}\delta \mu_i^\infty-e^{\nu_0/2}w_0{\pmb\nabla} \tilde{U}, 	\label{eu1}
\end{gather}
where the superscript $\infty$ denotes the corresponding redshifted quantity,
i.e. $\delta\mu_i^\infty\equiv e^{\nu_0/2}\delta\mu_i$. The derivation of this equation is straightforward and uses the following relations:
\begin{gather}
\frac{1}{w_0}{\pmb\nabla}P_0=-{\pmb\nabla}\frac{\nu_0}{2},
\quad
\frac{\p\mu_k}{\p n_i}=\frac{\p\mu_i}{\p n_k},
\quad
\delta n_i{\pmb\nabla} \mu_{i\,0}^\infty=0.
\end{gather}
The first relation represents the condition of hydrostatic equilibrium and can be derived from the equilibrium Euler equation. The second relation follows from thermodynamic relations \eqref{tdrel}. The third relation follows from the matter quasineutrality ($e_i n_i=0$) and combined conditions of chemical and hydrostatic equilibrium of the unperturbed matter ($e_k \mu_{i0}^\infty-e_i\mu_{k0}^\infty=\const$) \cite{kgk2024}:
\begin{gather}
\delta n_i{\pmb\nabla} \mu_{i0}^\infty=\delta n_i{\pmb\nabla}[\mu_{i0}^\infty-(e_i/e_k)\mu_{k0}^\infty]=0,
\end{gather}
where the index $k$ can refer to any charged particle species of the fluid (no summation over $k$ is assumed).

Now, multiplying Eq.~(\ref{eu1}) by $\dot {\bs \xi}$ and performing integration over the stellar volume $dV\equiv e^{\lambda_0/2}r^2 \sin \theta\,  drd\theta d\varphi$, we obtain
\begin{gather}
\frac{d}{dt}\int\frac{1}{2}e^{-\nu_0/2}w_0\dot{{\bs\xi}}^2 dV=A_U -\int n_{i0}(\dot{\bs \xi}\cdot{\pmb\nabla})\delta \mu_i^\infty dV,  \label{Ebal}
\end{gather}
where
\begin{gather}
A_U\equiv-\int e^{\nu_0/2}w_0
(\dot{\bs\xi}\cdot{\pmb\nabla})\tilde U\,dV
\end{gather}
is the power supplied to the star by the tidal force.
Next, integrating the second term on the right-hand side of Eq.~\eqref{Ebal} by
parts and neglecting the boundary term, which vanishes at the stellar surface
because $n_{i0}=0$ there, we find
\begin{multline}
\frac{d}{dt}\int\frac{1}{2}e^{-\nu_0/2}w_0\dot{{\bs\xi}}^2 dV= \int{\rm div}(n_{i0}\dot{\bs \xi})\,\delta \mu_i^\infty dV+A_U= \\
=\int e^{\nu_0}\Delta\Gamma_i \delta \mu_i dV-\int e^{\nu_0/2}\delta \dot n_i\delta \mu_i dV+A_U,
\end{multline}
where, in the last equality, we have used Eq.~\eqref{cont} to exclude ${\rm div}(n_{i0}\dot{\bs \xi})$. Then, based on the symmetry relation $\partial \mu_k/\partial n_i=\partial \mu_i/\partial n_k$ and the expression \eqref{deltaPmu} for the chemical potential perturbations, we can rewrite the second term on the right-hand side as 
\begin{gather}
\hspace{-0.2cm}\int e^{\nu_0/2}\delta \dot n_i\delta \mu_i dV=\frac{d}{dt}\int\frac{1}{2}e^{\nu_0/2}\delta n_i\biggl(\frac{\partial \mu_i}{\partial n_k}\biggr)_0\delta n_k dV,
\end{gather}
and, therefore, obtain
\begin{multline}
\frac{d}{dt}\int\left[\frac{1}{2}e^{-\nu_0/2}w_0\dot{{\bs\xi}}^2 +\frac{1}{2}e^{\nu_0/2}\delta n_i\biggl(\frac{\partial \mu_i}{\partial n_k}\biggr)_0\delta n_k \right]dV=\\
=\int e^{\nu_0}\Delta\Gamma_i \delta \mu_i dV+A_U.
\end{multline}

This equation has a rather clear physical interpretation of the energy conservation law. Specifically, it states that the change $dE_{\rm mech}/dt$ in the mechanical energy of the perturbation $E_{\rm mech}$ (left-hand side) is caused by the work $A_U$ of external gravitational forces and dissipative energy losses $\dot{E}_{\rm diss}$ due to out-of-equilibrium chemical reactions:
\begin{gather}
	\label{Ebalance}
	\frac{d E_{\rm mech}}{dt}=\dot{E}_{\rm diss}+A_U, \\
    E_{\rm mech}\equiv \int\left[\frac{1}{2}e^{-\nu_0/2}w_0\dot{{\bs\xi}}^2 +\frac{1}{2}e^{\nu_0/2}\delta n_i\delta\mu_i\right]dV, \label{Emech} \\
    \label{Edotdiss1}
    \dot{E}_{\rm diss}\equiv\int e^{\nu_0}\Delta\Gamma_i \delta \mu_i dV.
\end{gather}
The interpretation of $\dot{E}_{\rm diss}$ as describing the energy change due
to chemical reactions is straightforward. The quantity $\delta\mu_i$ is the
energy required to add one particle of species $i$ to the
system. Hence, each term $\Delta\Gamma_i\delta\mu_i$ represents the rate, per
unit volume, at which the energy changes due to the production or
destruction of particles of this species. After summation over all species and
integration over the stellar volume, this gives Eq.~\eqref{Edotdiss1}.
It may not be immediately obvious, however, why this contribution is
dissipative, i.e. why it is always non-positive,
$\dot E_{\rm diss}\leq 0$. To show this, it is useful to introduce the
following notation. Consider a reaction pair
$(a)=(12\leftrightarrow 34)$, with the direct reaction defined as
$12\to 34$. Let $N_{k(a)}$ be the number of particles of species $k$
produced in one direct reaction. Thus, for the reaction $12\to34$, one has
$N_{1(a)}=N_{2(a)}=-1$ and $N_{3(a)}=N_{4(a)}=+1$.

We also define
\begin{gather}
    \Delta\Gamma_{(a)}
    =
    \Gamma_{12\to34}-\Gamma_{34\to12},
\end{gather}
where $\Gamma_{12\to34}$ and $\Gamma_{34\to12}$ are the rates of the direct
and inverse reactions, respectively. Finally, we introduce the chemical
imbalance
\begin{gather}
\label{mua}
    \mu_{(a)}
    =
    -\mu_k N_{k(a)}
    =
    \mu_1+\mu_2-\mu_3-\mu_4 .
\end{gather}
With this convention, $\mu_{(a)}$ is the energy released in one direct
reaction $12\to34$. A positive value of $\mu_{(a)}$ therefore means that the
direct reaction is energetically favored over the inverse one. For
thermodynamically consistent reaction rates, this implies that
$\Delta\Gamma_{(a)}$ has the same sign as $\mu_{(a)}$. Conversely, if
$\mu_{(a)}<0$, the inverse reaction is favored and
$\Delta\Gamma_{(a)}<0$. In chemical equilibrium,
$\mu_{(a)0}=0$ and $\Delta\Gamma_{(a)0}=0$, so that, to linear order,
$\delta\mu_{(a)}=\mu_{(a)}$.

We can now prove that $\dot E_{\rm diss}$ is non-positive. The particle
production rates can be written as
\begin{gather}
    \Delta\Gamma_i=N_{i(a)}\Delta\Gamma_{(a)},
\end{gather}
where summation over reaction pairs $(a)$ is implied. Substituting this
expression into Eq.~\eqref{Edotdiss1} and using the definition
\eqref{mua}, we obtain
\begin{gather}
\label{EdotdissNeg}
    \dot E_{\rm diss}
    =
    -\int e^{\nu_0}
    \Delta\Gamma_{(a)}\delta\mu_{(a)}\,dV
    \leq 0 .
\end{gather}
The last inequality follows because
$\Delta\Gamma_{(a)}\delta\mu_{(a)}\geq0$ for each reaction pair.

For the particular chemical composition and set of chemical reactions we consider, the sum over all reactions $(a)$ in the formula above can be equivalently rewritten as
\begin{gather}
\label{representation}
\delta\mu_{(a)}\Delta\Gamma_{(a)}=\delta \mu\,\Delta \Gamma+\delta \mu_{\rm s}\,\Delta \Gamma_{\rm s}, \\
\delta \mu\equiv \mu_n-\mu_\Lambda, \qquad \delta \mu_{\rm s}\equiv 2\mu_\Lambda -(\mu_{\Xi^-}+\mu_p).
\end{gather}
The quantity $\delta \mu_{\rm s}$ represents the chemical potential imbalance \eqref{mua} associated with the strong process \eqref{strong}, while $\delta\mu$ is the chemical imbalance associated with any weak nonleptonic process \eqref{reac1}-\eqref{reac3}. Next, $\Delta\Gamma$ denotes the production rate of $\Lambda$-hyperons or, equivalently, the consumption rate of neutrons in the weak nonleptonic processes \eqref{reac1}-\eqref{reac3}. Finally, $\Delta \Gamma_{\rm s}$ represents the production rate of protons/$\Xi^-$-hyperons or, equivalently, half the consumption rate of $\Lambda$-hyperons, associated with the strong process \eqref{strong}. 

Using representation \eqref{representation}, we rewrite the energy dissipation rate $\dot{E}_{\rm diss}$ \eqref{EdotdissNeg} as
\begin{gather}
\dot{E}_{\rm diss}\equiv-\int (\delta \mu\,\Delta \Gamma+\delta \mu_{\rm s}\,\Delta \Gamma_{\rm s})\, e^{\nu_0}dV. \label{Ediss}
\end{gather}
The structure of $\dot{E}_{\rm diss}$ suggests the existence of two
non-dissipative limiting regimes: the frozen-reaction regime (extremely slow reactions) and the
instantaneous-equilibrium regime (extremely fast reactions). In the former case, the particle source terms
are small while the chemical potential imbalances remain finite. In the latter
case, the particle source terms may remain finite, but the chemical potential
imbalances are efficiently relaxed. In both limits, the product
$\Delta\Gamma_{(a)}\delta\mu_{(a)}$
vanishes, and so does the corresponding
dissipation.
Consequently, within the approximations adopted here, the contribution
$\delta\mu_{\rm s}\Delta\Gamma_{\rm s}$ to $\dot{E}_{\rm diss}$ vanishes,
because the strong process is treated as equilibrated:
$\delta\mu_{\rm s}\to0$, while $\Delta\Gamma_{\rm s}$ remains finite.

Generally, $\Delta \Gamma$ is a nonlinear function of $\delta \mu$. In the so-called subthermal regime, however, when $|\delta\mu|<T$, one can approximate $\Delta\Gamma$ by the corresponding leading order (linear) term in $\delta\mu$, and write:
\begin{gather}
\label{lambdadef}
\Delta \Gamma=\lambda \delta \mu,
\end{gather}
where the coefficient $\lambda$ represents the so-called total reaction rate associated with the weak processes \eqref{reac1}-\eqref{reac3}%
\footnote{In what follows, we assume that the inequality $|\delta\mu|<T$ holds and verify it numerically in Sec.~\ref{num}.}.
In this limit, the hydrodynamic equations can be equivalently reformulated in terms of an effective bulk viscosity $\zeta$ and chemically modified adiabatic index $\gamma$. The transition from one description to another relies on separating the adiabatic part of the pressure perturbation from the dissipative one: $\delta P= \delta P_{\rm ad}+\delta P_{\rm diss}$. The explicit form of $\delta P_{\rm ad}$ and $\delta P_{\rm diss}$ is uniquely determined from continuity equations \eqref{cont}. Specifically, using linearized thermodynamic relations \eqref{tdrel} and continuity equations \eqref{cont}, 
one expresses $\delta P$ and $\delta\varepsilon$ as linear functionals
of the Lagrangian displacement $\bs{\xi}$ and then identifies dissipative and nondissipative terms. 
Eventually, one finds that $\delta\varepsilon$ does not contain a dissipative contribution (see, e.g., \cite{kgk24,kgk25}), 
and 
$\delta P_{\rm diss}$, $\delta P_{\rm ad}$, and $\delta \varepsilon$
can be obtained from the equations
\begin{gather}
\label{PdissDef}
\delta P_{\rm diss}=-\zeta e^{-\nu_0/2}\divv\dot{\bs{\xi}}, \\
\label{PadDef}
\Delta P_{\rm ad}=\frac{\gamma P_0}{w_0}\Delta\varepsilon=-\gamma P_0 \divv\bs{\xi},
\end{gather}
where $\Delta f\equiv\delta f+(\bs{\xi}\cdot \bs{\nabla})f_0$ denotes the Lagrangian perturbation of a scalar $f$, and the information about the chemical reactions is encoded in the effective bulk viscosity coefficient $\zeta$ and chemically modified adiabatic index $\gamma$. In general, these coefficients depend on the density of the matter, its temperature, and also on the perturbation frequency. Their explicit form is automatically established when deriving \eqref{PdissDef}-\eqref{PadDef}, once the relevant microphysics is specified. 
For the chemical composition and reactions considered in this study, a detailed
derivation of $\gamma$ and $\zeta$ can be found in \cite{kgk25}. 
The same
reference also discusses in detail the conditions under which the slow- and
fast-reaction limits are valid.

The Navier–Stokes equation in the discussed effective description takes the following form (see, e.g., \cite{ll87,gusakov07,gkcg13,kgk25})
\begin{multline}
e^{-\nu_0}w_0\ddot{\bs\xi}=-{\pmb\nabla}\delta P_{\rm ad}+\frac{\delta P_{\rm ad}+\delta\varepsilon}{w_0}{\pmb\nabla} P_0-w_0{\pmb\nabla} \tilde{U} + \\
+e^{-\nu_0/2}{\bs \nabla}(\zeta \,{\rm div}\dot{\bs \xi}), \label{ns}
\end{multline}
where the last term arises from $\delta P_{\rm diss}$ and accounts for the dissipative effect of reactions [compare with Eq.\ \eqref{eu}]. It has the same form as in the case of dissipation due to kinetic bulk viscosity, that is usually described by introducing the additional term
\begin{gather}
T^{\mu\nu}_{\rm bulk}\equiv-\zeta[g^{\mu\nu}+u^\mu u^\nu](\nabla_\rho u^\rho)
\end{gather}
to the perfect-fluid stress-energy tensor $T^{\mu\nu}$ \eqref{perfectTmunu}.

One can show that the energy dissipation rate \eqref{EdotdissNeg} expressed in terms of the effective bulk viscosity is given by \cite{ll87,gusakov07,gkcg13,kgk25}
\begin{gather} 
    \dot{E}_{\rm diss}=-\int \zeta \left(\nabla_\mu u^\mu\right)^2 e^{\nu_0} dV=-\int \zeta ({\rm div} \dot{\bs\xi})^2  dV. \label{Edissxi}
\end{gather}
This formula can be derived in a way very similar to the derivation of Eq.\ \eqref{Edotdiss1}: via integrating the Euler equation \eqref{ns} multiplied by $e^{\nu_0/2}\dot{\bs{\xi}}$.

As we mentioned previously, one of the effects of dissipation is the heating of the neutron star’s hyperon core at a rate $(-\dot{E}_{\rm diss})$. Assuming, for simplicity, that the redshifted temperature of the hyperon core, $T^\infty$, is uniform (although this is generally not the case%
\footnote{The thermal conductivity timescale exceeds the inspiral timescale, so the temperature profile should evolve with time, depending on the local rate of energy dissipation and heat capacity. 
})
while the temperature of the outer layers remains unchanged, we find that $T^\infty$ evolves in time according to 
\begin{gather}
	\label{heatEq}
	C\frac{dT^\infty}{dt}=-\dot{E}_{\rm diss},
\end{gather}
where $C$ is the heat capacity of the hyperon stellar core. 

Another important effect caused by dissipation is the decrease of the orbital energy $E_{\rm orb}$ of the binary system, which accelerates the inspiral compared to the adiabatic case, when $E_{\rm orb}$ decreases solely due to emission of GW. This acceleration is conveniently quantified by the corresponding shift $\Delta \phi_{\rm diss}$ that dissipation introduces to the phase of the gravitational-wave signal. Assuming that the shift accumulates between the times $t_1$ and $t_2$, it can be approximately calculated using the following formula (for the derivation see, e.g., \cite{kgk24}):
\begin{gather}
	\label{phaseShift}
	\Delta \phi_{\rm diss}\approx -\int\limits_{t_1}^{t_2} 2\Omega(t) \frac{\dot E_{\rm diss}}{\dot E_{\rm orb}} \, dt,
\end{gather}
where $\Omega(t)$ is the frequency of the orbital motion. Within the adopted perturbative approach, when evaluating the phase shift, one can use functions $\Omega(t)$ and $\dot E_{\rm orb}$, calculated within the point-mass approximation (see Introduction), since any corrections due to finite-size effects would produce higher order contributions to $\Delta \phi_{\rm diss}$.

Note that all other effects contributing to the phase shift (see, e.g., Refs.\ \cite{dhs21,kgk24}), which are neglected here, are cumulative. In particular, this includes the adiabatic effect of reactions on the tidal deformability, arising from the dependence of $\gamma$ on the reaction rate \cite{ap20}.

\subsection{Estimating $\dot{E}_{\rm diss}$}\label{sec24}

To determine the phase shift $\Delta\phi_{\rm diss}$ or the stellar heating,
we first need to find the Lagrangian displacement $\bs{\xi}$ that enters
Eq.~\eqref{Edissxi} for $\dot{E}_{\rm diss}$. To this end, we adopt a
perturbative approach: we determine $\bs{\xi}$ by neglecting dissipation, i.e.,
as a solution of the adiabatic Euler equation \eqref{ns} with $\zeta=0$,
subject to the boundary conditions discussed at the end of Sec.~\ref{perturb},
namely regularity at the stellar center and vanishing total pressure at the
perturbed stellar surface. For this purpose, it is convenient to substitute
explicitly $\delta P_{\rm ad}$ and $\delta\varepsilon$, expressed as linear
functionals of the Lagrangian displacement according to Eq.~\eqref{PadDef}.
Once the pressure and energy-density perturbations are expressed through
$\bs{\xi}$, the Euler equation \eqref{ns} in the adiabatic limit
($\zeta\to0$) takes the following form:
\begin{gather}
	\label{EulerEqNoDiff}
	e^{-\nu_0/2}w_0\ddot{\bs\xi}-\hat{\mathcal{L}}_{\rm ad}{\bs\xi}=-e^{\nu_0/2}w_0{\pmb\nabla}\tilde{U},
\end{gather}
with the operator $\hat{\mathcal{L}}_{\rm ad}{\bs\xi}$ defined as
\begin{widetext}
\begin{gather}
	\hat{\mathcal{L}}_{\rm ad}{\bs\xi}\equiv -e^{\nu_0/2}\, {\pmb\nabla}\delta P_{\rm ad}+e^{\nu_0/2}\,\frac{\delta P_{\rm ad}+\delta\varepsilon}{w_0}{\pmb\nabla} P_0={\pmb \nabla}\left[
	\left(
	{\pmb \nabla}P_0 \cdot {\bs \xi} + \gamma P_0 \, {\rm div}{\bs \xi}
	\right) {e}^{\nu_0/2}
	\right] 
	+\left(
	{\pmb \nabla} \varepsilon_0 \cdot {\bs \xi} + w_0 \, {\rm div} {\bs \xi}
	\right) {\pmb \nabla}(e^{\nu_0/2}).
\end{gather}
\end{widetext}
Once the solution $\bs{\xi}$ is found, it can be used in the expression \eqref{Edissxi} to estimate the leading order contribution to $\dot{E}_{\rm diss}$. 

In what follows, we consider two limits: when the reactions are slow and when they are fast. As discussed above, in these limits the dissipation is, indeed, small, and the dissipative term only weakly affects $\bs \xi$. Moreover, calculations of the dissipation rate of freely oscillating hyperon stars (e.g., \cite{gk08,kgk25}) indicate that the dissipative term in Navier-Stokes equation remains small even beyond these limits, which justifies our approach across the entire parameter range. 

To proceed further, we need to specify the explicit form of the gravitational
tidal potential $\tilde U$ generated by the companion star. In this study, we model the companion star as a point-mass object. Let us choose the orientation of the coordinate system such that $\theta=\pi/2$ corresponds to the equatorial plane of the binary orbit (recall that the origin of the coordinate system is placed at the center of mass of the perturbed neutron star). Then, in the $(r,\theta,\varphi)$-coordinates, the components of the companion star position vector $\bs{D}(t)$ are ${\bs D}(t)=\{D(t),\pi/2,\Phi(t)\}$, and the full gravitational potential $U$ induced by the companion star inside the neutron star is given by
\begin{multline}
	U=-\frac{GM'}{|{\bs r}-{\bs D}(t)|}=\\
    =-GM'\sum_{lm}W_{lm}\frac{r^l}{D(t)^{l+1}}{e}^{-i m \Phi(t)}Y_{lm}(\theta,\varphi). \label{Uexp}
\end{multline}
Here, $G$ is the gravitational constant, $M'$ is the mass of the companion, $Y_{lm}$ are the spherical harmonics and $W_{lm}$ are known numerical coefficients (see, e.g., Ref.~\cite{Lai94}). The $l=m=0$ term in this expansion does not depend on $\bs r$ and, therefore, does not produce any acceleration. The sum of all the $l=1$ terms is given by $U_{l=1}=-(G M'/D^3)({\bs D}\cdot{\bs r})$ and results in a uniform acceleration. Therefore, by definition, $l=0$ and $l=1$ terms should be attributed to $U_0$ in the decomposition $U=U_0+\tilde{U}$ [recall the discussion before Eq.~\eqref{gtensor}] and the tidal potential is then given by
\begin{gather}
	\tilde{U}=-GM'\sum_{l\geq 2, m}W_{lm}\frac{r^l}{D(t)^{l+1}}{e}^{-i m \Phi(t)}Y_{lm}(\theta,\varphi). \label{Utilde}
\end{gather}
Since $r/D(t)\ll 1$, the main contribution to $\tilde{U}$ comes from the lowest, $l=2$, harmonics with $m=\{0,\pm 2\}$ (modes with $m=\pm 1$ do not contribute because $W_{21}=W_{2-1}=0$). Therefore, in what follows, we retain only the $l=2$ and $m=\{0,\pm 2\}$ contributions to the tidal potential $\tilde{U}$.  

To calculate ${\bs \xi}$, we expand the Lagrangian displacement into the basis $\{{\bs\xi}_{\bf k}\}$ of complex eigenfunctions ${\bs\xi}_{\bf k}$ of the operator $\hat{\mathcal{L}}_{\rm ad}$, where the index $\bf k$ labels different eigenfunctions (see also Refs.\ \cite{pt77,rg94,Lai94,kgk24}):
\begin{gather}
	\label{basis}
	{\bs\xi}(t,{\bs r})=\sum_{\bf k}a_{\bf k}(t){\bs \xi}_{\bf k}({\bs r}),
    \\
    \label{eigenvalue_problem}
	\left\{
	\begin{gathered}
        \hat{\mathcal{L}}_{\rm ad}{\bs\xi}_{\bf k}+e^{-\nu_0/2}w_0\omega_{\bf k}^2{\bs\xi}_{\bf k}=0, \\
		\Delta P_{{\rm ad}\,\bf k}|_{r=R}=0, \\
		{\bs\xi}_{\bf k} \text{ is regular at $r=0$},
	\end{gathered}
	\right.
\end{gather}
where $R$ is the stellar radius.
The numbers $\omega_{\bf k}$ represent the eigenfrequencies of the freely oscillating neutron star, and vectors ${\bs\xi}_{\bf k}$ describe the spatial dependence of the corresponding eigenfunctions (note that, in this basis, the boundary conditions for ${\bs\xi}$ are satisfied automatically). In what follows, for convenience, we normalize these functions by the following condition:
\begin{gather}
	\label{orthonorm}
	\int e^{-\nu_0/2}w_0 \, {\bs\xi}^\star_{\bf k}\cdot{\bs\xi}_{\bf k'} dV=\delta_{\bf k k'}.
\end{gather}

To find the $\bf k$-mode amplitude $a_{\bf k}(t)$, we expand $\bs\xi$ into the introduced basis, multiply the Euler equation~\eqref{EulerEqNoDiff} by ${\bs\xi}_{\bf k}^\star$, integrate over the stellar volume, and, using the orthogonality \eqref{orthonorm} of the basis, arrive at the following equation for the coefficient $a_{\bf k}$:
\begin{gather}
	\label{EulerEqak}
	\ddot a_{\bf k} + \omega_{\bf k}^2 a_{\bf k}=\frac{GM' 
    W_{lm}
    Q_{\bf k}}{D(t)^{l+1}}{e}^{-i m \Phi(t)},
    \\
    Q_{\bf k}=\int dV {e}^{\nu_0/2}w_0\pmb \, {\pmb \xi}_{\bf k}^\star \cdot \pmb \nabla[r^l 
    Y_{lm}
    (\theta,\varphi)].
\end{gather}
where $m$ is the magnetic quantum number of the mode ${\bf k}$. To proceed further, we, following Ref.~\cite{Lai94}, introduce the function $b_{\bf k}(t)$:
\begin{gather}
	\label{bkint}
	a_{\bf k}(t)=GM'
    W_{lm}
    Q_{\bf k} b_{\bf k}(t)e^{-i m \Phi(t)}. 
\end{gather}
Defined this way, the functions $b_{\bf k}(t)$ evolve on the timescale $\sim D(t)/\dot{D}(t)$ of the orbital evolution caused by GW emission. Substituting Eq.~(\ref{bkint}) into Eq.~\eqref{EulerEqak}, we then obtain
\begin{gather}
	\label{EulerEqbk}
	\ddot b_{\bf k} -2 i m \Omega \dot b_{\bf k}+(\omega_{\bf k}^2-m^2\Omega^2-i m \dot\Omega) b_{\bf k}=\frac{1}{D(t)^{l+1}},
\end{gather}
where we have replaced $\dot{\Phi}=\Omega$. As long as $m\Omega\ll \omega_{\bf k}$, which holds at 
sufficiently large separation distances $D$, the asymptotic solution obtained by neglecting the $d/dt$ terms provides a good approximation:
\begin{gather}
	\label{bkAsympt}
    b_{\bf k}(t)=\frac{1}{D^{l+1}(t)[\omega_{\bf k}^2-m^2\Omega^2(t)]}. 
\end{gather}
Further, one can use ${\bs\xi}$, defined by 
Eqs.~\eqref{basis}, \eqref{bkint}, and \eqref{bkAsympt},
to calculate the dissipation rate $\dot{E}_{\rm diss}$ given by Eq.~(\ref{Edissxi}). Noting that $\bs\xi$ is real-valued (even though the basis eigenfunctions are complex), we have
\begin{multline}
	\label{Edissak}
	\dot{E}_{\rm diss}=-\int 
    \zeta\left({\rm div}\,\dot{\bs\xi}\right)^2
    dV=-\int \zeta{{\rm div}\dot{\bs\xi}}\,{\rm div}{\dot{\bs\xi}^\star} dV=\\
    =-\sum_{\bf kk'}\dot a_{\bf k}(t)\dot a_{\bf k'}^\star(t)\int \zeta{{\rm div}{\bs\xi_{\bf k}}}{\rm div}{{\bs\xi}^\star_{\bf k'}} dV, 
\end{multline}
Note that the cross terms in Eq.~\eqref{Edissak} do not vanish unless the modes have different angular dependence (e.g., \cite{kgk24}).

Strictly speaking, the accurate calculation of $\dot{E}_{\rm diss}$ requires summation over all modes $\bf k$ and $\bf k'$. 
Such a calculation is rather involved; below we simplify it by adopting several approximations.
First of all, we note that the $m=0$ modes actually almost do not contribute to $\dot E_{\rm diss}$, since they perturb the stellar matter on the slow timescale of the binary evolution [see Eq.\ \eqref{bkint} with $m=0$]. Dissipation caused by such slow perturbations is negligibly weak, since they cannot significantly drive weak nonleptonic processes out of equilibrium, which, therefore, remain equilibrated [recall the discussion that accompanies Eq.\ \eqref{Ediss}].
In contrast, perturbations with $m=\pm 2$ occur much faster, approximately at the frequency $\omega=2\Omega(t)$. Therefore, we suggest simplifying the calculation by keeping only the $l=2$ and $m=\pm 2$ modes.

Next, it is well known (see, e.g., \cite{Lai94,ap20}) that the leading contribution to $\bs{\xi}$ comes from the so-called f-modes, and it is tempting to retain only 
f-modes in the decomposition of $\bs{\xi}$ \eqref{basis}. By doing this one can show that
\begin{gather}
\divv\bs{\xi}|_\text{f-mode}\sim 0.1\,\frac{\xi}{R}, \label{fscaling}
\end{gather}
where $\xi$ is a typical magnitude of $\bs{\xi}$, and the factor $0.1$ arises numerically due to well-known weak compressibility of the f-modes. This estimate, however, substantially differs from the result known from the treatment not resorting to eigenmode decomposition. 
Specifically, such calculation reveals that, in nonbarotropic matter at low perturbation frequency $\omega=|m|\Omega=2\Omega$, $\divv\bs{\xi}$ is suppressed by a factor $\sim(\omega/\mathcal{N})^2\sim (2\Omega)^2/\omega_{\rm g}^2$, where $\mathcal{N}$ is the so-called Brunt-V$\ddot{\rm a}$is$\ddot{\rm a}$l$\ddot{\rm a}$ frequency, and $\omega_{\rm g}\sim \mathcal{N}$ is the eigenfrequency of the main $l=2$ g-mode harmonic [$\omega_{\rm g}/(2\pi)\approx 567\,\rm Hz$ in our numerical model], see, e.g., \cite{weinberg16} (their equation 62) and \cite{aw19,div}
%
\footnote{In this respect, dissipation due to diffusion at early stages of the inspiral proposed in \cite{kgk24} is likely inefficient. The analysis of \cite{kgk24} did not account for the suppression of $\rm{div} {\bs\xi}$.}:
%
\begin{gather}
{\rm div}{\bs\xi}\sim\frac{\omega^2}{\mathcal{N}^2}\frac{\xi}{R}\sim \frac{(2\Omega)^2}{\omega_{\rm g}^2}\frac{\xi}{R}.
\label{scaling}
\end{gather}
We verified this scaling by performing low-frequency response calculations for a simplified stellar model and for external potential frequencies of $1\,\rm Hz$ and higher. 
Away from the resonances with high-order g-modes, our numerical results confirm the suppression given by Eq.~\eqref{scaling}. Eq.~\eqref{scaling} shows that an f-mode-only truncation would overestimate the compressive
part of the tide. In the full nonbarotropic response, the f-mode contribution
to ${\rm div}\pmb{\xi}$ is largely cancelled by contributions from
the remaining modes, yielding the 
scaling above.
Dissipation driven by $\rm{div} {\bs\xi}$ (e.g., bulk viscosity or diffusion) does not alter the situation. It becomes evident from the consideration of a limiting case of static tide, when $\divv\bs{\xi}=0$. In this case, the solution to the dissipative equations coincides with that of the non-dissipative ones, as the dissipative term vanishes for this solution. Based on the order-of-magnitude estimates above, and combining Eqs.~\eqref{fscaling} and \eqref{scaling}, we see that
\begin{gather}
\divv\bs{\xi}\sim 10\frac{(2\Omega)^2}{\omega_{\rm g}^2}\divv\bs{\xi}|_\text{f-mode}.
\end{gather}
To account approximately for this suppression of $\divv\bs{\xi}$ when
calculating $\dot{E}_{\rm diss}$, we propose replacing the summation over all
modes in Eq.~\eqref{Edissak} by a summation over the f-modes only, multiplied
by a phenomenological reduction factor $\mathcal{R}(\Omega)$:
\begin{gather}
\sum_{(\bf{k,k'})\in\text{all modes}}\to \mathcal{R}\sum_{(\bf{k,k'})\in\text{f-modes}},
\end{gather}
where $\mathcal{R}(\Omega)$ is given by
\begin{gather}
\label{red}
\mathcal{R}(\Omega)\equiv
\left\{
\begin{gathered}
100 \frac{(2\Omega)^4}{\omega_{\rm g}^4}, \hspace{0.2cm} 100 \frac{(2\Omega)^4}{\omega_{\rm g}^4}\leq 1, \hfill \\
1, \hspace{1.5cm} 100 \frac{(2\Omega)^4}{\omega_{\rm g}^4}> 1. \hfill
\end{gathered}
\right.
\end{gather}
Then, using this simplified approach, the energy dissipation rate \eqref{Edissak} can be estimated by the following formula:
\begin{multline}
	\label{Edissak1}
	\dot{E}_{\rm diss}\approx-\mathcal{R}\sum_{\bf k}\dot a_{\bf k}(t)\dot a_{\bf k}^\star(t)\int \zeta{{\rm div}{\bs\xi_{\bf k}}}{\rm div}{{\bs\xi}^\star_{\bf k}} dV \equiv \\
	\equiv \mathcal{R}\sum_{\bf k}\dot a_{\bf k}(t)\dot a_{\bf k}^\star(t)\dot{E}_{\rm diss\, \bf k},
\end{multline}
where the summation runs solely over the $l=2$, $m=\pm 2$ f-modes.

\begin{figure}[t]
	\center{\includegraphics[width=0.75\linewidth]{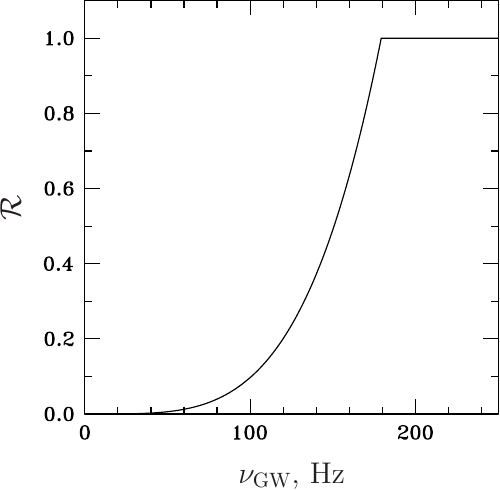}}
	\caption{The reduction factor $\mathcal{R}$ as a function of the emitted gravitational wave frequency.}
	\label{Fig:Red}
\end{figure}

\section{Results of numerical calculations}
\label{num}

We now turn to quantifying the impact of nonequilibrium particle transformations
during the inspiral.
In our numerical calculations, we construct a hybrid stellar model. We use the
FSU2H equation of state (EOS) \cite{pfpr19} for the stellar core and the
analytic fit to the BSk24 EOS \cite{pcp19} for the crust. The crust-core
interface is placed at the density at which the FSU2H and BSk24 pressures
coincide, $\rho=2.45\times10^{14}\,{\rm g\,cm^{-3}}$.
As a reference configuration, we consider a general relativistic neutron star model with central density $\rho_{\rm c}\approx 10^{15}\,\rm g\,cm^{-3}$, radius $R\approx 1.3\times 10^6\,\rm cm$, and mass  $M=1.85M_\odot$%
\footnote{
Although neutron stars as massive as $1.85M_\odot$ have not yet been observed
in Galactic double neutron star systems, such systems cannot be ruled out.
For instance, a neutron star with a mass of 
$1.65M_\odot$ (PSR J1913+1102) 
has already been detected \cite{ferdman18,fzt19}. Moreover, the mass distributions of Galactic double neutron stars and merging double neutron stars may differ \cite{Kruckow20}. The high total mass of the binary system that produced GW190425 also supports this possibility \cite{rs20,galaudage21}.
}.
Such a neutron star contains both $\Lambda$- and $\Xi^-$-hyperons in its inner core. The hyperon core is relatively extended, with radius and enclosed mass approximately equal to $0.56R$ and $0.34M$, respectively. Concerning the bulk viscosity coefficient $\zeta$, it is calculated following \cite{oghf19}. 
Specifically, we use equation~(13) and the fitting formulas (15), (41), and (42)
of Ref.~\cite{oghf19}.
A correction factor of $1.4$ is applied to the parameter $\zeta_0$ entering equation (15) of \cite{oghf19}, as described at the end of their section~V. Finally, the heat capacity $C$ of the hyperon core is calculated following \cite{yls99}, assuming that all particle species contribute to $C$ except for the strongly superconducting $\Xi^-$ hyperons.

To estimate the stellar heating and the GW phase shift caused by chemical
reactions, we first calculate $\dot E_{\rm diss}$.
In the subthermal regime, which, as verified below, holds during most of the inspiral, the dissipation rate $\dot E_{\rm diss}$ can be approximated by Eq.~\eqref{Edissak1} (see also \cite{cls90,gkcg13}). 
All calculations below are performed for a companion mass
$M'=1.4\,M_\odot$.
We use Eq.~\eqref{Edissak1} with the asymptotic solutions of the mode amplitudes $a_{\bf k}$, computed adopting Eqs.~\eqref{bkint} and \eqref{bkAsympt}. The required eigenmodes $\bs{\xi}_{\bf k}$ and their frequencies $\omega_{\bf k}$ are determined as solutions of the eigenvalue problem \eqref{eigenvalue_problem}. The time-dependence of the orbital frequency $\Omega(t)$ that enters these equations is calculated within the point-mass approximation. Below, we express the evolution of the system (in particular, the stellar heating and accumulated phase shift) in terms of  
the frequency $\nu_{\rm GW}=2\Omega/2\pi$ of the emitted gravitational waves,
rather than the time $t$.

In general, the eigenfunctions $\bs\xi_{\bf k}$ appearing in the expression for $\dot E_{\rm diss}$ depend on the rate of nonleptonic reactions through the adiabatic index $\gamma$. Once the particle sources $\Delta\Gamma_k$ are specified, the continuity equations \eqref{cont} with thermodynamic relations \eqref{thermo1}, \eqref{deltaPmu}, and \eqref{tdrel} allow one to determine $\gamma$ at arbitrary reaction rates, as was done in \cite{kgk25}. Here, however, we do not follow this rigorous approach and instead employ a simplified treatment. Specifically, we evaluate $\bs\xi_{\bf k}$ in two limiting cases: (i) when the nonleptonic reactions \eqref{reac1}-\eqref{reac3} are slow and thus almost do not alter the composition, and (ii) when they are fast, so that the stellar matter remains nearly in equilibrium with respect to these reactions. 
We then calculate the dissipation rate
$\dot E_{\rm diss \, \bf k}$ in these two limits and interpolate between them, as described below.

To determine the appropriate interpolation, we first note that the transition between the slow- and fast-reaction regimes is governed locally by the dimensionless ratio $\tau_{\rm eq}^{-1}/\omega$ \cite{kgk25}, where $\omega$ is the perturbation frequency  (in our case, $\omega \approx 2\Omega$) and $\tau_{\rm eq}^{-1}\equiv\mathcal{B}\lambda$ is the chemical equilibration rate with $\mathcal{B}$ being the appropriate thermodynamic susceptibility (in the notation of Ref.~[6], $\mathcal{B}=n_b^{-1}|\partial\delta\mu/\partial y_s|$). This rate is often denoted by $\gamma$ in the bulk-viscosity literature (see, e.g., \cite{ah21}); we use $\tau_{\rm eq}^{-1}$ here to avoid confusion with the chemically modified adiabatic index $\gamma$. The slow- and fast-reaction limits correspond to $\tau_{\rm eq}^{-1}\ll\omega$ and $\tau_{\rm eq}^{-1}\gg\omega$, respectively.  In other words, the eigenmodes obtained as solutions of the eigenvalue problem \eqref{eigenvalue_problem} with the chemically modified adiabatic index $\gamma$ depend on the total reaction rate $\lambda$ only through the ratio $\tau_{\rm eq}^{-1}/\omega$.

In degenerate matter, $\mathcal{B}$ is temperature-independent, while the weak nonleptonic reaction rate satisfies $\lambda\propto T^2\propto(T^\infty)^2$  (see, e.g., \cite{oghf19}). Thus, the transition can equivalently be parametrized by the dimensional variable
\begin{gather}
v \equiv (T^\infty/10^8\,{\rm K})^2/\omega= (T^\infty/10^8\,{\rm K})^2/(2\Omega),
\end{gather}
where $T^\infty$ and $T$ are the redshifted and local stellar temperatures, respectively. Note that, since the stellar matter is treated as degenerate, the temperature dependence of the eigenmodes is caused solely by that of the total reaction rate $\lambda$ and, therefore, can also be entirely described in terms of the transition parameter $v$.

Further, we note that the dependence of the bulk viscosity coefficient $\zeta$ on $T^\infty$ and $\omega$ can be parametrized as $\omega^{-1}F(v)$, where  $F(v)$ is some function of the parameter $v$ (see equation 13 in \cite{oghf19}).  
Consequently, keeping in mind that the temperature dependence of the eigenmodes is also parametrized by $v$, we can write 
$\dot E_{\rm diss\,\bf k}\propto \omega^{-1} \tilde{F}(v)$,
where $\tilde F(v)$ is a 
function of $v$.
This result suggests that the quantity 
$\omega\dot E_{\rm diss\,\bf k}\propto \tilde{F}(v)$
can be considered as a function of a single parameter $v$. 
We therefore evaluate 
$\omega\dot E_{{\rm diss},{\bf k}}$
for a range of $v$ values in the slow- and fast-reaction limits and fit the result with
\begin{align}
\label{fits}
\omega\dot E_{\rm diss\,\bf k}=-k_1 v/(1+k_2 v^2),
\end{align} 
which smoothly interpolates between the two limits%
\footnote{One can show \cite{kgk25} that in these limits the eigenmodes weakly depend on $v$, and the $v$-dependence of $\dot{E}_{\rm diss\,\bf k}$ is determined by that of bulk viscosity, whose asymptotic behavior in the considered limits matches that of the suggested interpolation formula.}.
Here, $k_1$ and $k_2$ are fitting parameters 
specific for each mode $\bf k$. 
For each of the $l=2$, $m=\pm 2$ f-modes we found $k_1=1.94\times 10^{54}\,\rm erg/sec$ and $k_2=4.23\times 10^7\,\rm sec^{-2}$ [note, that $\dot E_{\rm diss\,\bf k}$ has units $\rm erg\, sec$, see Eq.~\eqref{Edissak1}].
Notably, the values of eigenfrequencies and overlap integrals almost coincide in the two limits and are equal to $\omega_{\bf k}\approx 1.37\times 10^{4}\,\rm s^{-1}$ and $Q_{\bf k}\approx 0.36$ in our numerical model. 
Using the resulting fits, together with the explicit expressions for the mode amplitudes
$a_{\bf k}$ and the reduction factor $\mathcal{R}$ \eqref{red}, we determine with Eq.~\eqref{Edissak1}
the energy dissipation rate $\dot{E}_{\rm diss}=\dot{E}_{\rm diss}(\Omega,T^\infty)$ as a function of the redshifted temperature $T^\infty$ and orbital frequency $\Omega(t)$.

\begin{figure}[t]
	\center{\includegraphics[width=0.75\linewidth]{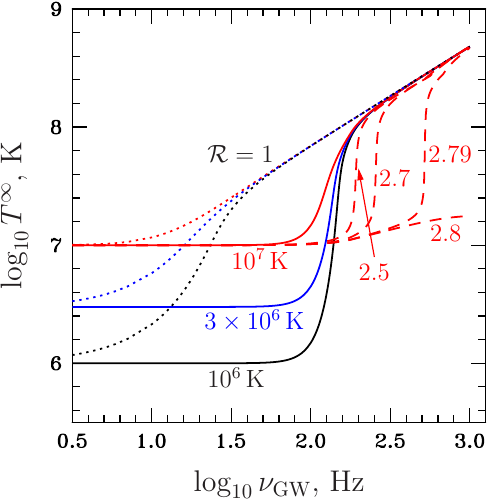}}
	\caption{Redshifted hyperon-core temperature as a function of
$\nu_{\rm GW}$ for the three initial stellar temperatures:
$10^6\,{\rm K}$, $3\times 10^6\,{\rm K}$, and $10^7\,{\rm K}$ (solid lines). Dotted lines show the corresponding results for $\mathcal{R}=1$, while dashed lines show the effect of neutron superfluidity. Numbers next to the dashed curves indicate the assumed uniform redshifted neutron critical temperature, normalized to $10^8\,\rm K$. See text for details.}
	\label{Fig:T}
\end{figure}

We now compute the stellar heating and the GW phase shift.
To this end, we replace $\dot E_{\rm diss}$ with its approximate form given by Eq.~\eqref{Edissak1}, where, as announced above, we use analytic fits \eqref{fits} for $\dot E_{\rm diss \,\bf k}$ and adopt $\omega(t)=2\Omega(t)$. To calculate the reduction factor $\mathcal{R}$ entering Eq.~(\ref{Edissak1}), we 
determine $\omega_{\rm g}$, and evaluate Eq.~(\ref{red})
\footnote{To evaluate $\omega_{\rm g}$ we work in the limit of slow reactions \eqref{reac1}-\eqref{reac3}, since, as Fig.~\ref{Fig:Red} suggests, $\mathcal{R}$ differs from unity at rather early stages of the inspiral, when hyperon-core temperature is still sufficiently low (see Fig.\ \ref{Fig:T}), so that reactions \eqref{reac1}-\eqref{reac3} are slow. 
As we mentioned earlier, $\omega_{\rm g}/(2\pi)\approx 567\,\rm Hz.$}
. The resulting dependence of the reduction factor $\mathcal{R}$ on the frequency $\nu_{\rm GW}$ of the emitted gravitational waves is shown in Fig.~\ref{Fig:Red}. Once the function $\dot{E}_{\rm diss}(\Omega,T^\infty)$ is determined, we can proceed to finding the temperature evolution $T^\infty(t)$. We specify initial stellar temperature at a large orbital separation, where dissipative heating does not yet affect the thermal state (at $\nu_{\rm GW}=0.1\,\rm Hz$ in our numerical calculations). Solving Eq.~\eqref{heatEq} then gives the time evolution of the redshifted temperature of the hyperon core, $T^\infty(t)$.
This, in turn, allows us to calculate the phase shift $\Delta \phi_{\rm diss}$ by substituting the approximation \eqref{Edissak1} for $\dot E_{\rm diss}(t)\equiv\dot E_{\rm diss}(\Omega(t),T^\infty(t))$ into Eq.~\eqref{phaseShift}. Since the functions $T^\infty(t)$ and $\Omega(t)$ are already known, Eq.~(\ref{phaseShift}) can be readily integrated.

The resulting dependence of the redshifted temperature $T^\infty$ on the frequency $\nu_{\rm GW}$ of the emitted gravitational waves is shown in Fig.~\ref{Fig:T} by solid lines for three initial values of $T^\infty$ (at $\nu_{\rm GW}=0.1\,\rm Hz$): $T^\infty=10^6\,\rm K$, $T^\infty=3\times 10^6\,\rm K$, and  $T^\infty=10^7\,\rm K$. Dotted and dashed lines are discussed at the end of this section.
At the early stages of the inspiral ($10\,\rm Hz\la \nu_{\rm GW}\la 100\,\rm Hz$), the temperature is still low, so that the system is in the slow-reaction
regime. During this stage, $T^\infty$ increases exponentially with $\nu_{\rm GW}^{16/3}$
and remains sensitive to its initial value.
After the transition to the fast-reaction regime, which occurs at
$\nu_{\rm GW}\sim150\,{\rm Hz}$, $T^\infty$ becomes insensitive to the
initial temperature and follows the scaling $T^\infty\propto \nu_{\rm GW}^{5/6}$ [the reduction factor in Eq.\ \eqref{Edissak1} equals unity at this late stage of the inspiral, see Fig.\ \ref{Fig:Red}]. 
During the inspiral, the stellar temperature reaches several times
$10^8\,{\rm K}$, significantly higher than the typical values
$\sim 10^7\,{\rm K}$ obtained in previous calculations
\cite{Lai94,yw17,kksk22,kgk24}. 

\begin{figure}[t]
	\center{\includegraphics[width=0.75\linewidth]{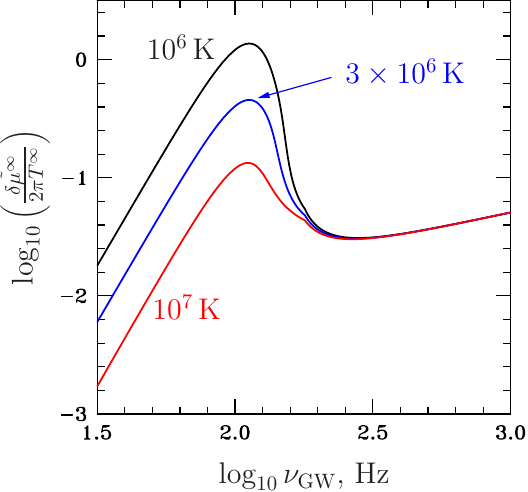}}
	\caption{Ratio $\tilde{\delta \mu^\infty}/(2\pi T^\infty)$ as a function of the gravitational-wave frequency $\nu_{\rm GW}$ for three initial stellar temperatures: $10^6\,\rm K$, $3\times 10^6\,\rm K$, and $10^7\,\rm K$ (labeled in the plot). See text for details.}
	\label{Fig:subthermal}
\end{figure}

Our calculations are valid only in the subthermal regime, where $\delta \mu \ll T$. 
To verify that the inspiral indeed proceeds in this regime, we introduce the averaged redshifted chemical imbalance, defined as
\begin{gather}
\tilde{\delta \mu^\infty}\equiv\left(-\dot{E}_{\rm diss}/I_\lambda\right)^{1/2},\;\;
I_\lambda=\int \lambda dV.
\end{gather} 
This definition of $\tilde{\delta \mu^\infty}$ is motivated by the general formula \eqref{Ediss} for the energy dissipation, valid beyond the subthermal regime, and the definition \eqref{lambdadef} of the total reaction rate $\lambda$, introduced in the subthermal regime.
In Fig.~\ref{Fig:subthermal}, we plot $\tilde{\delta \mu^\infty}/(2\pi T^\infty)$
as a function of $\nu_{\rm GW}$%
\footnote{In general, $\Delta \Gamma =\lambda \delta \mu\left[1+\delta \mu^2/(2\pi T)^2\right]$, see equation (40) of \cite{oghf19} and \cite{hly02}. Thus the true smallness parameter is $\delta \mu/(2\pi T)$.}.
The fact that this quantity remains below unity for almost the entire inspiral%
\footnote{It slightly exceeds unity at the early stage of the inspiral in our example with the initial temperature $T^\infty=10^6\,\rm K$. 
A more accurate treatment beyond the subthermal regime would lead to somewhat faster heating at $100\,\rm Hz\la \nu_{\rm GW}\la 150\,\rm Hz$, but this is not expected to significantly affect our results.}
confirms the validity of our assumption that the inspiral proceeds in the subthermal regime.

\begin{figure}[t]
	\center{\includegraphics[width=0.75\linewidth]{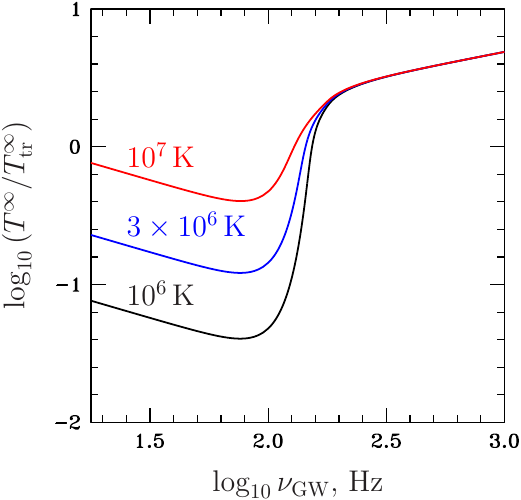}}
	\caption{The ratio of the stellar hyperon core temperature to the temperature $T_{\rm tr}$ (at which the bulk viscosity reaches its maximum efficiency) as a function of $\nu_{\rm GW}$ for three initial stellar temperatures, $10^6\,\rm K$, $3\times 10^6\,\rm K$, and $10^7\,\rm K$ (labeled in the plot). See text for details.}
	\label{Fig:TTmax}
\end{figure}

At first glance, this result may seem to contradict Ref.~\cite{aw19}, which considered
NSs composed of neutrons, protons, and leptons. That reference found that the chemical
potential imbalance greatly exceeds the stellar temperature during the inspiral.
In fact, however, there is no contradiction: hyperon nonleptonic reactions are much
faster than the Urca processes considered in Ref.~\cite{aw19} and thus efficiently
relax the chemical potential imbalances, keeping the system in the subthermal regime.
This is illustrated in Fig.\ \ref{Fig:TTmax}, where we show the ratio of the stellar hyperon core temperature to the transition temperature $T_{\rm tr}$ defined as the temperature corresponding to the maximum of $|\dot E_{\rm diss}|$
at a given perturbation frequency [the maximum arises due to nonmonotonic behavior of the bulk viscosity $\zeta$ as a function of the total reaction rate $\lambda$, see, e.g., \cite{kgk25}, and in our numerical implementation corresponds to $T_{\rm tr}=10^8 \sqrt{2\Omega}/k_2^{1/4}$~K, see Eq.\ \eqref{fits}]. 
At temperatures well below $T_{\rm tr}$, oscillations proceed with an almost
frozen composition, so the chemical imbalances generated by the perturbations
do not relax efficiently (as in \cite{aw19}), while at temperatures well above $T_{\rm tr}$, the chemical composition remains nearly equilibrated throughout the oscillation cycle. 
Fig.\ \ref{Fig:TTmax} indicates that, at the final stages of the inspiral, when strong departures from chemical equilibrium due to stellar deformations could in principle occur, the system actually resides in the latter regime. This resolves the apparent discrepancy with \cite{aw19}.

\begin{figure}[t]
	\center{\includegraphics[width=0.75\linewidth]{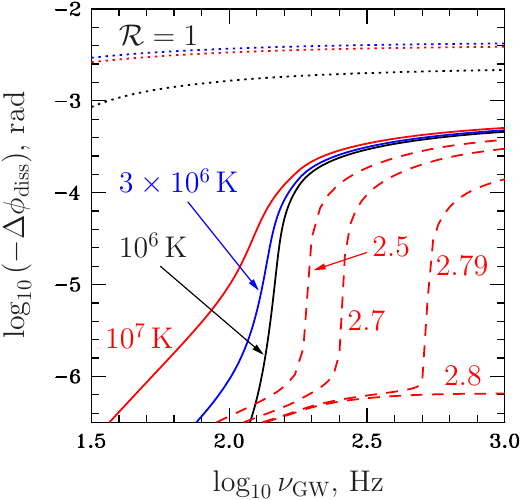}}
	\caption{Phase shift due to bulk viscosity as a function of $\nu_{\rm GW}$. Notations are the same as in Fig.~\ref{Fig:T}.}
	\label{Fig:phase}
\end{figure}

The phase shift $\Delta\phi_{\rm diss}(\nu_{\rm GW})$, calculated using
Eq.~\eqref{phaseShift}, is shown in Fig.~\ref{Fig:phase}. Solid lines correspond to the three initial temperatures, discussed above. Other curves are discussed at the end of this section.
The arbitrary additive constant in the phase shift is chosen so that
$\Delta\phi_{\rm diss}(\nu_{\rm GW}=1\,{\rm Hz})=0$.
The figure shows that the bulk-viscosity
contribution to the phase is $\lesssim10^{-3}\,{\rm rad}$, well below the
sensitivity of the LIGO and Virgo detectors. Moreover, because this contribution
is degenerate with other physical effects, it is unlikely to be measured and
disentangled observationally. These findings are consistent with the estimates
of Ref.~\cite{most22}, but are in strong tension with the recent results of
Ref.~\cite{gpkd24}, who reported much higher stellar temperatures and
significantly larger phase shifts for inspiraling hyperonic stars.

Further, in order to test the sensitivity of our results to the adopted assumptions, we performed additional calculations. First, we considered the case $\mathcal{R}=1$ to illustrate the role of the reduction factor $\mathcal{R}$. This choice deliberately removes the low-frequency suppression of $\mathrm{div}\boldsymbol{\xi}$ discussed in Sec.~\ref{sec24}.
The corresponding results, shown by dotted lines in Figs.~\ref{Fig:T} and \ref{Fig:phase} for three initial temperatures, demonstrate that increasing $\mathcal{R}$ allows dissipation to become more efficient at earlier stages of the inspiral and, consequently, leads to a larger accumulated phase shift. Nevertheless, even for $\mathcal{R}=1$, the resulting phase shift remains at the level of only a few times $10^{-3}$~rad.

In addition, we carried out calculations for the case where, besides strongly superconducting $\Xi^-$ hyperons, neutrons are also moderately superfluid. In this case, we accounted for the suppression of reactions \eqref{reac1}--\eqref{reac3} using the superfluid reduction factors calculated in Ref.~\cite{kgk25}, included the suppression of the neutron heat capacity using the fit from Ref.~\cite{yls99}, and modified the fitting formula \eqref{fits}. The results of these calculations are shown in Figs.~\ref{Fig:T} and \ref{Fig:phase} by dashed lines. Next to each curve, we indicate the redshifted critical neutron temperature, in units of $10^8 \,\mathrm{K}$, assumed to be constant throughout the hyperonic core. The initial temperature was taken to be $10^7 \,\mathrm{K}$. As seen from the figures, neutron superfluidity delays the onset of efficient dissipation and, consequently, reduces the accumulated phase shift. In turn, proton superconductivity, if present, would further delay the onset of efficient dissipation and would therefore reduce the accumulated phase shift even more. It should be noted, however, that in these calculations we assumed $\delta\mu \ll T$, which, strictly speaking, is not satisfied throughout the inspiral in the presence of neutron superfluidity. A fully self-consistent treatment would lead to less delayed dissipation, followed by the temperature approaching an asymptotic value. We leave such calculations beyond the scope of the present work.

Finally, to assess the maximum possible impact of hyperon bulk viscosity within the present framework, we performed an additional conservative calculation. In this calculation, we treated neutrons as nonsuperfluid, set $\mathcal{R}=1$ in Eq.~\eqref{Edissak1}, and adopted the maximum absolute value of $\dot E_{{\rm diss},{\bf k}}$ at each orbital frequency $\omega=2\Omega(t)$, corresponding to $T=T_{\rm tr}$. Thus, the calculation simultaneously removes the low-frequency reduction of $\mathrm{div}\boldsymbol{\xi}$ and maximizes the bulk-viscous dissipation at every stage of the inspiral. The resulting phase shift may therefore be regarded as an upper-limit estimate within our model. Integrating Eq.~\eqref{phaseShift} under these assumptions yields
$\Delta\phi_{\rm diss}=-7.7\times10^{-3}\, {\rm rad}$ at $\nu_{\rm GW}=1000\, {\rm Hz}$ for our reference
$1.85\,M_\odot$~---~$1.4\,M_\odot$ binary.

Notably, we find that our upper-limit estimate is much smaller than the phase shift reported in Ref.~\cite{gpkd24}. To examine whether the mass asymmetry is responsible for the discrepancy with Ref.~\cite{gpkd24}, we repeated the described maximally conservative calculation for an equal-mass $1.85\,M_\odot$~---~$1.85\,M_\odot$ binary, accounting for bulk-viscous dissipation in both stars. In this case we obtain $\Delta\phi_{\rm diss}\simeq-1.0\times 10^{-2}\, {\rm rad}$ at $\nu_{\rm GW}=1000\, {\rm Hz}$. Thus, removing the mass asymmetry changes the upper-limit estimate only moderately and cannot account for the substantially larger phase shifts reported in Ref.~\cite{gpkd24}.
Likewise, the difference between the bulk viscosity coefficients used in the two calculations is also insufficient to explain the discrepancy fully. At $n_b=2.5n_0$ and matched perturbation frequencies, the peak values of the bulk viscosity inferred from Fig.~3 of
Ref.~\cite{gpkd24} are approximately a factor of $2$--$3$ larger than the corresponding values obtained with the FSU2H-based viscosity prescription adopted here. Although this difference may contribute to the discrepancy, it is still not pronounced enough to explain it completely.
Unfortunately, a definitive comparison of the two calculations is hindered by the fact that some details
required to reconstruct the dissipation rate of Ref.~\cite{gpkd24}
are not specified.

\section{Conclusions}
\label{disc}

During binary neutron star inspiral, tidal forces drive the stellar matter out of mechanical and chemical equilibrium. The resulting nonequilibrium perturbations can be damped by various dissipative mechanisms, which convert part of the orbital energy into heat inside the star. In this study, we examined the effect of particle reactions in hyperonic neutron-star matter, usually described in terms of hyperon bulk viscosity.
Contrary to what might be expected from the estimates of Ref.~\cite{gpkd24}, we find that hyperon bulk viscosity has only a very small effect on the gravitational-wave phase of inspiraling neutron stars. 
In our models, the corresponding phase shift does not exceed a few times $10^{-3}$~rad (see Fig.~\ref{Fig:phase}), even though the adopted stellar model possesses a relatively extended hyperon core.
Importantly, this small phase shift is not a consequence of the reduction
factor \eqref{red} introduced to model the low-frequency suppression of
$\mathrm{div}\boldsymbol{\xi}$: even a maximally conservative calculation with
$\mathcal{R}=1$ gives $|\Delta\phi_{\rm diss}|\lesssim 10^{-2}\,\mathrm{rad}$ up to
$\nu_{\rm GW}=1000\,\mathrm{Hz}$.
This is far below the sensitivity of currently operating gravitational-wave detectors and is likely too small to be disentangled even with next-generation observatories, especially in view of other theoretical and astrophysical uncertainties.
At the same time, hyperon bulk viscosity can noticeably affect the thermal evolution of the hyperonic core. In particular, in the models considered here, viscous heating can raise the core temperature up to approximately $5\times 10^8\,{\rm K}$; see Fig.~\ref{Fig:T}.

\section*{Acknowledgements}
The groundwork for this study was laid during a long-term visit by the authors to the Weizmann Institute of Science (WIS). We acknowledge the support of the visit by the Simons Foundation and WIS. The authors are grateful to the Department of Particle Physics \& Astrophysics at WIS for their hospitality and excellent working conditions. The authors were supported by RSF [Grant № 22-12-00048-P].


\label{lastpage}

\end{document}